\documentclass{article}
\usepackage{mathrsfs}
\usepackage{pifont}
\usepackage{graphicx}
\usepackage{amsmath,amssymb,amsthm}

\newcommand{\ini}{\begin{equation}}
\newcommand{\fin}{\end{equation}}
\newcommand{\inia}{\begin{eqnarray}}
\newcommand{\fina}{\end{eqnarray}}

\begin{document}

\title{\bf de Sitter and double irregular domain walls }

\author{Rommel Guerrero \footnote{grommel@ucla.edu.ve}, R. Omar Rodriguez \footnote{romar@uicm.ucla.edu.ve},\\
Rafael Torrealba\footnote{rtorre@uicm.ucla.edu.ve}, R. Ortiz\footnote{rortiz@uicm.ucla.edu.ve}}

\date{ \it Unidad de Investigaci\'on en Ciencias Matem\'aticas, Universidad
Centroccidental Lisandro Alvarado, 400 Barquisimeto, Venezuela}
\maketitle
\vspace{0.3cm}

\begin{abstract}
A new method to obtain thick domain wall solutions to the coupled
Einstein scalar field system is presented.
 The procedure allows the construction of irregular walls from well known ones, such that
the spacetime associated to them are physically different. As
consequence of the approach, we obtain two irregular geometries
corresponding to thick domain walls with $dS$ expansion and
topological double kink embedded in $AdS$ spacetime. In
particular, the double brane can be derived from a fake
superpotential.
\end{abstract}
\vspace{0.3cm}
\hspace*{25pt}\noindent KEY WORDS: thick branes, de Sitter walls, BPS walls

\vspace{0.3cm}

\section{Introduction}
The gravitational properties of domain walls or 3-branes have been
object of intense investigation for different reasons. On one
hand, it has been pointed out that four-dimensional gravity can be
realized on a thin wall interpolating between Anti-de Sitter ($AdS$) spacetimes
\cite{Randall:1999ee,Randall:1999vf,Cvetic:1993xe}. On the other hand, wall configurations are
relevant for the study of Renormalization-Group flow Equations in the context of Anti-de Sitter/Conformal Field Theory correspondence \cite{Maldacena:1997re}.

Considering our four-dimensional universe as an infinitely thin
domain wall is an idealization, and it is for this reason, that in
more realistic models the thickness of the brane has been taken
into account
\cite{DeWolfe:1999cp,Gremm:1999pj,Csaki:2000fc,Kehagias:2000au,Kehagias:2000dg,Behrndt:2001km,Kobayashi:2001jd,Campos:2001pr}.
The thick domain walls are solutions to Einstein's gravity theory
interacting with a scalar field, where the scalar field is a
standard topological kink interpolating between the minima of a
potential with spontaneously broken symmetry. The nonlinearity of
the five-dimensional Einstein's gravity theory interacting with a
scalar field represents a strong difficulty when finding exact
solutions to the system. In this sense, several formulations and
methodologies
\cite{Skenderis:1999mm,DeWolfe:1999cp,Guerrero:2002ki,Bazeia:2002xg,Melfo:2002wd}
have been reported, where it is possible to reconsider the problem
in a simpler way. Maintaining this focus, in this paper, we
develop a new formulation or method based on the linearization of
one of the equations of the system.

We show that the coupled system with planar symmetry, admits two
solutions, physically not equivalent, associated to a scalar
field, with different self-interaction potentials. We have called
these solutions regular and irregular solutions, because in the
last case the geometry spacetime hosts a coordinate singularity
defined by a hypersuperface of codimension one. In
\cite{Guerrero:2005aw} we made simultaneous use of the regular and
irregular solutions and showed that it is possible to generate
asymmetric brane worlds with interesting properties, where the
irregular contribution plays an important role on the asymmetry in
the brane. In this sense, in the present paper, we analyze the
explicit realization of the domain walls on those irregular
spacetimes.

In section III,  we consider a regular brane with a de Sitter ($dS$) expansion
\cite{Goetz:1990} and we find that the irregular geometry
corresponds to a domain wall with similar features interpolating
between two $AdS$ vacua. It is well known that due to the
non-linearity and instability of the gravitational interactions,
the inclusion of the gravitational evolution into a dynamic thick
wall is a highly non-trivial problem. For this reason, there are
not so many analytic solutions of a dynamic thick domain wall. In
fact, to our knowledge, in the literature  encountered so far,
there only exist two solutions \cite{Goetz:1990,Sasakura:2002tq},
with background metric on the brane given by a $dS$ expansion,
which resembles the Friedmann-Robertson-Walker metric, typical in
a cosmological framework.

In section IV, we obtain an irregular topological double kink
geometry embedded into an $AdS$ background, from the double
regular configuration analyzed in \cite{Melfo:2002wd}, and show
that this irregular spacetime is derivable from a fake
superpotential. In the context of static domain wall spacetime,
the double walls are topological defects richer than the standard
topological kinks. These configurations may be seen as defects
that host internal structure, representing two parallel walls,
whose properties have been studied in several papers: in
\cite{Campos:1998db}, using models described by a complex scalar
field; in \cite{Morris:1995zp,Bazeia:1995ui,Edelstein:1997ej}, in
models described by two real scalar fields on flat spacetime; and
in \cite{Gregory:2001xu}, in brane scenarios involving two higher
dimensions on curved spacetime. On the other hand, the models
supported on a single real scalar field, with a $2$-kink profile
have only been considered in \cite{Melfo:2002wd,Bazeia:2003aw},
thus the irregular double domain wall considered here, corresponds
to another example of these exotic configurations.

Finally, in section V we summarize our results.

\section{Approach to generate new solutions}
Consider the action of Einstein's gravity theory interacting with a real scalar field $\phi$ 
\inia
S=\int d^{4}y dx \sqrt{- \mathrm{g}}\left[\frac{1}{2}R-\frac{1}{2}\mathrm{g}^{\mu\nu}\nabla_\mu\phi\nabla_\nu\phi-V(\phi)\right]\label{Accion},
\fina
where $\mathrm{g}$ is the metric tensor, $R$ is the scalar curvature and $V(\phi)$ the scalar potential. The field equations generated from last action has the form  
 \inia
 &&G_{ab}=R_{ab}-\frac{1}{2}\ \mathrm{g}_{ab}\ R=T_{ab},\label{Einstein-scalar0}\\ 
 &&T_{ab}=\nabla_a\phi\ \nabla_b\phi-\mathrm{g}_{ab}\left[\frac{1}{2}\ \nabla^c\phi\
 \nabla_c\phi+V(\phi)\right],\\ 
 && \nabla^c \nabla_c\phi-\frac{d V(\phi)}{d\phi}=0, \label{Einstein-scalar}
 \fina
with $G_{ab}$ the Einstein tensor, $R_{ab}$ the Ricci tensor and $T_{ab}$ the stress-energy tensor.

Let $\mathrm{g}$ be the metric tensor for a 5-dimensional spacetime with planar-paralell symmetry, given by \cite{Taub}
 \ini
 \mathrm{g}_{ab}=f(x)^2\ \left(-dt_a\ dt_b+e^{2\beta t}\ dy_a^j\
 dy_b^j\right)+f(x)^2\ dx_a\ dx_b,\quad\beta>0 ,\label{MetricGeneral}
 \fin
where $\ j=1,2,3\ $ and $a,\ b=0,\dots ,4$. We look for solutions
to the coupled Einstein-scalar field system (\ref{Einstein-scalar0}-\ref{Einstein-scalar}) satisfying the requirements
 \begin{description}
 \item[\bf{1.-}] $\ \phi=\phi(x)$,
 \item[\bf{2.-}] $\lim_{x\rightarrow\pm\infty}\phi(x)=\varphi_{\pm}$\:\: such that\:\: $\lim_{\phi\rightarrow\varphi_{\pm}} \frac{d V(\phi)}{d\phi}=0$,\:\:\: $\varphi_{\pm}\in\Re$
 \item[\bf{3.-}] $\ \phi^{\prime}(x)^2$ is symmetric under reflections in the
 $x=0$ plane,
 \end{description}
\begin{flushleft}where the prime denotes derivative with respect to $\ x$.
Following the usual strategy, we find
 \inia
 \phi^{\prime}(x)^2&=&f^2\ \left(G_4{}^4-G_0{}^0\right)=6\ \frac{f^{\prime\
 2}}{f^2}-3\ \frac{f^{\prime\prime
}}{f}-3\ \beta^2\ \label{EcPhi},\\
V(\phi(x))&=&-\frac{1}{2}\ \left(G_4{}^4+G_0{}^0\right)=-3\
\frac{f^{\prime\
 2}}{f^4}-\frac{3}{2}\ \frac{f^{\prime\prime
}}{f^3}+\frac{9}{2}\frac{1}{f^2}\ \beta^2.
 \fina
\end{flushleft}

Now, if we rewrite (\ref{EcPhi}) in terms of the inverse function
of the metric factor, i.e. $g=1/f$, we obtain
 \ini
 g^{\prime\prime}(x)-\left[\frac{1}{3}\
 \phi^{\prime}(x)^2+\beta^2\right]\ g(x)=0.
 \fin
This equation is the Sturm-Liouville type and it is well known
that has an orthogonal set of solutions $\ \{g_1,\ g_2\}\ $
associated, where
 \ini
 g_2(x)=g_1(x)\ \chi(x);\qquad \chi(x)\equiv\int_{x_0}^x\
 \frac{1}{g_1(\xi)^2}\ d\xi \label{g2}
 \fin
and the general solution
 \ini
 g(x)=g_1(x)\left(c_1+c_2\ \chi(x)\right),\qquad c_1,\ c_2 \in \Re \label{GeneralSol}
 \fin
depends on two arbitrary constans.

Remarkably enough, equation (\ref{g2}) gives a mechanism to obtain
new solutions to the coupled system from known ones, such as
$g_{1}$, compatible with the same scalar field but with different
potentials. We are interested in the realization of domain walls
on the geometry defined by $g$, in such sense we will consider
$g_{1}$ as functions whose analytic behavior generate new domain
wall solutions to (\ref{Einstein-scalar0}-\ref{Einstein-scalar}).

Let $\ g_{1}\ $ be a function such that
\begin{flushleft}
 \begin{description}
 \item[\bf{1.-}] $\ g_1\ $ is $C^{2}$,
 \item[\bf{2.-}] $\ g_1\ >\ 0$ (positive definite),
 \item[\bf{3.-}] $\ g_1\ $ is a function asymptotically
 increasing,\newline
 \end{description}
then $f_1$ is a continuous, integrable and asymptotically
vanishing function, which we will call \emph{regular}; and $g$ is
smooth function with a zero in some value of $x$, corresponding to
a singularity in $f$, for the following cases
\begin{description}
\item[1.-] If $c_{1}, c_{2}\neq 0$, then $g$ will have a zero in $x_{p}\in x$ when $\chi(x) |_{x_{p}}=-c_{1}/c_{2}.$
\item[2.-] If $c_{1}=0$ and $c_{2}\neq 0$, then $g$ will have a zero in $x_{0}$.
\end{description}
\end{flushleft}

In the first case, it is possible to avoid the divergence in
(\ref{MetricGeneral}). It is only necessary to choose
appropriately the constants $c_{1}$ and $c_{2}$, so that its
negative quotient does not belong to the image of $\chi(x)$. In
the ref. \cite{Guerrero:2005aw}, we developed and discussed this
kind of solutions for two exotic branes and show that it is
possible to localize the four-gravity on them. In the second case
the divergence in (\ref{MetricGeneral}) is unavoidable, however in
this paper we explore the possibility to set up domain walls on
this geometry, where the metric factor is singular or
\emph{irregular}.

To study the above irregular solutions further, let us first note
that all the scalars built from the Riemann tensor are finite in
the whole bulk. Thus, the spacetime described by the irregular
solutions are free of scalar singularities. Moreover, for the
hypersurfaces $\ x=x_{0}$, the tidal forces experimented by a
freely falling observer become bounded. To show this explicitly,
let us consider the nonvanishing components of the Riemann
curvature tensor with respect to the static orthonormal frame
$R_{0j0j}=-R_{1k1k}=-R_{2323}$, $R_{j4j4}=-R_{0404}$ where
$j=1,2,3;\ k=2,3$ and
 \ini
R_{2323}=\beta^2\ g^2-g^{\prime 2}\ ,\qquad R_{0404}=g^{\prime
2}-g\ g^{\prime\prime}\ .
 \fin
Free falling observers with energy $E$ are connected to the static
orthonormal frame by a local Lorentz boost in the direction
perpendicular to the wall, with an instantaneous velocity given by
 \ini
 v^{\hat{4}}=\sqrt{1-E^{-2}\ g^{-2}}\ .
 \fin
Then the nonvanishing curvature components in the Lorentz boosted
frame are $R_{\hat{k}\hat{4}\hat{k}\hat{4}}$,
$R_{\hat{2}\hat{3}\hat{2}\hat{3}}$,
$R_{\hat{0}\hat{k}\hat{0}\hat{k}}$,
$R_{\hat{0}\hat{4}\hat{0}\hat{4}}$,
$R_{\hat{0}\hat{k}\hat{k}\hat{4}}=R_{\hat{k}\hat{4}\hat{0}\hat{k}}$
where $\hat{k}=\hat{2},\hat{3}$ and
 \inia
&& R_{\hat{k}\hat{4}\hat{k}\hat{4}}=-E^2\ g^3\ (\beta^2\ g-g^{\prime\prime})-g^{\prime 2}, \qquad\qquad\quad \label{Tidal1} \\ 
&& R_{\hat{0}\hat{k}\hat{0}\hat{k}}=-E^2\ g^3\ (\beta^2\ g-g^{\prime\prime})+g^{\prime 2}-g\ g^{\prime\prime},\\
&& R_{\hat{0}\hat{k}\hat{k}\hat{4}}=-E\ g\ \sqrt{E^2\ g^2-1}\ (\beta\ g-g^{\prime\prime}),\\
&&R_{\hat{2}\hat{3}\hat{2}\hat{3}}=\beta^2\ g^2-g^{\prime 2},\\
&&R_{\hat{0}\hat{4}\hat{0}\hat{4}}=g^{\prime 2}-g\ g^{\prime\prime}\label{Tidal2}
\fina

According to equations (\ref{Tidal1}-\ref{Tidal2}), none of the
components of Riemann in the explorer's orthonormal frame become
infinite in the pathological point $x=x_0$. Therefore, the metric
$\mathrm{g}$ on the surface $x=x_0$ is a nonsingular and perfectly
well-behaved region of spacetime\footnote{In the static case similar conclusions can be obtain from the approach of the effective geometry \cite{Novello:2000km}.}. We think that there must be a
differentiable chart for which the resulting differentiable
structure gives a smooth metric, but to find these coordinates is
not the purpose of this paper.

As a final comment, it is important to remark that geometries
obtained from $g_{1}$ and $g$ with metric tensors $\mathrm{g}_{1\mu\nu}$
and $\mathrm{g}_{\mu\nu}$ are physically different, because it does not
exist a diffeomorphism among them. In fact, the spacetimes are
connected by a conformal transformation given by
\begin{equation}
\mathrm{g}_{1\mu\nu}=\Omega^{2}\mathrm{g}_{\mu\nu}, \qquad \Omega\equiv c_{1}+c_{2}\chi(x).
\end{equation}
In concordance with \cite{Wald:1984rg}, the conformal
transformations should not be understood as diffeomorphisms, but
as applications that allow to connect two different spacetimes
with identical causal structure. In this sense, we can conclude
that the manifolds with metric tensors $\mathrm{g}_{1\mu\nu}$ and
$\mathrm{g}_{\mu\nu}$ are physically not equivalent.

\section{Irregular domain wall with a de Sitter expansion}

Consider the embedding of a
thick $dS$ 3-brane into a five-dimensional bulk described by the
metric (\ref{MetricGeneral}), with a reciprocal metric factor
given by
 \begin{eqnarray}
g_{1}(x)=\cosh^{\delta}\left(\beta x/\delta \right),\quad
1/2>\delta>0,\quad \beta>0,\label{f goetz}
 \end{eqnarray}
In Ref. \cite{Goetz:1990,Gass:1999gk}, it has been shown that this spacetime is a solution to the
coupled Einstein-scalar field equations with
\begin{eqnarray}
\phi=\phi_{0}\arctan\left(\sinh\ \beta x/\delta\right),\qquad
\phi_{0} =\sqrt{3\delta(1-\delta)},\label{campo goetz}
\end{eqnarray}
and
\begin{eqnarray}
V_{1}(\phi)=\frac{1+3\delta}{2\delta}\ 3\beta^{2}\left(\cos
\phi/\phi_{0} \right)^{2(1-\delta)}.\label{potencial goetz}
\end{eqnarray}
The scalar field takes values $\ \pm\ \phi_{0}\ \pi/2\ $ at $\
x\rightarrow\pm\ \infty$, corresponding to two consecutive minima
of the potential with cosmological constant $\ \Lambda=0$, and
interpolates smoothly between these values at the origin; with $\
\delta\ $ playing the role of the wall's thickness. In \cite{Wang:2002pk} it has been shown that this domain
wall geometry localizes gravity on the wall, and in \cite{Guerrero:2002ki} that
it has a well-defined distributional thin wall limit\footnote{See Refs.\cite{Pantoja:2002nw,Pantoja:1997zt,Pantoja:2000cq} for more details on distributional description in General Relativity}.
We wish to find another thick domain with $dS$ expansion in five-dimensions. 
From equation (\ref{GeneralSol}),
choosing $c_{1}=0$ and $c_{2}=1$, with $\ g_1\ $ given by (\ref{f
goetz}); we obtain
 \begin{eqnarray}
g =\frac{i\ \delta}{\beta - 2\beta\delta}\
\cosh^{(1-\delta)}(\beta x/\delta)\ \csc\text{h}(\beta x/\delta)\ |\sinh\beta x/\delta|\\{}_2F_1[l, k, n, \cosh^2(\beta
x/\delta)],\qquad 1/2>\delta>0,\quad \beta>0,\label{newgeneral}
 \end{eqnarray}
 where ${}_{2}F_{1}$ is the hypergeometric function with $l=1/2-\delta,\ k=l+\delta$, and
 $n=l+1$. This solution represents a two-parameter family.
  Then, for simplicity we will consider the case $\delta=1/4$ without loss of generality
\begin{equation}
g(x)=-\frac{i}{2\ \beta}\ \cosh(2\ \beta x)^{1/4}\ F[2i\ \beta
x,\:\sqrt{2}],\qquad \beta>0,\label{g2dSIrregular}
\end{equation}
where $F$ is the incomplete first order elliptic function \cite{Arfken}. This
spacetime is solution to (\ref{Einstein-scalar}) with
 \begin{equation}
\phi=\phi_{0}\arctan\left(\sinh\ 4\beta x\right),\qquad
 \phi_{0}=3/4,\label{CampoGoetz2}
 \end{equation}
 and
\begin{eqnarray}
V(\phi)&=&-6 |\cos\ 4\phi/3|^{1/2}\nonumber \\ \nonumber
&& -\frac{21}{8} |\cos
4\phi/3|^{3/2} F[i/2 \arg\sinh\tan\ 4\phi/3
,2]^2\\ 
&&+|\cos\ 4\phi/3|\ \tan(4\phi/3)
F[i/2\ \arg\sinh\tan\ 4\phi/3 ,\:2],\label{PotGoetz2}
 \end{eqnarray}
where $\ \phi\ $ interpolates smoothly between the two degenerate
minima of $\ V(\phi)$, $\ \pm\ \phi_{0}\ \pi/2$. To this spacetime corresponds the following energy density
and pressure density
 \inia
\rho(x)&=&-\frac{3}{4}\ \cosh(4\ \beta x)^{-3/2}\left(8\ \cosh(4\ \beta x)
 +5\ F[2i\ \beta x,\:2]^2\right)\nonumber\\
 &&+6i\ \tanh(4\ \beta x)\ F[2i\ \beta x,\:2],
 \fina
 \inia
 p(x)&=&-\frac{3}{2}\ \cosh(4\ \beta x)^{-3/2}\left(4\ \cosh(4\ \beta x)
 +\ F[2i\ \beta x,\:2]^2\right)\nonumber\\
 &&+6i\ \tanh(4\ \beta x)\ F[2i\ \beta x,\:2].
  \fina

Notice that (\ref{g2dSIrregular}) has a zero  in $\ x= 0$. Hence,
the spacetime with tensor metric (\ref{MetricGeneral},
\ref{g2dSIrregular}) hosts a singularity defined by a hypersurface
of codimension one. However, the solution (\ref{g2dSIrregular},
\ref{CampoGoetz2}, \ref{PotGoetz2}) represents a one-parameter
family of plane symmetric irregular domain walls with a $dS$
expansion and reflection symmetry along the direction
perpendicular to the walls, being asymptotically $AdS_5$ with a
cosmological constant $\ \Lambda=-6\ \varepsilon$, where
$\varepsilon$ is the bound of
Image$\{\chi(x)\}=\{-\varepsilon/2\beta,\ +\varepsilon/2\beta\}$
and for $\ \delta=1/4$, $\ \varepsilon=1.31103\ $.

\begin {figure}[h]
\begin{minipage}[h]{0.45\linewidth}
\includegraphics[width=5cm,angle=0]{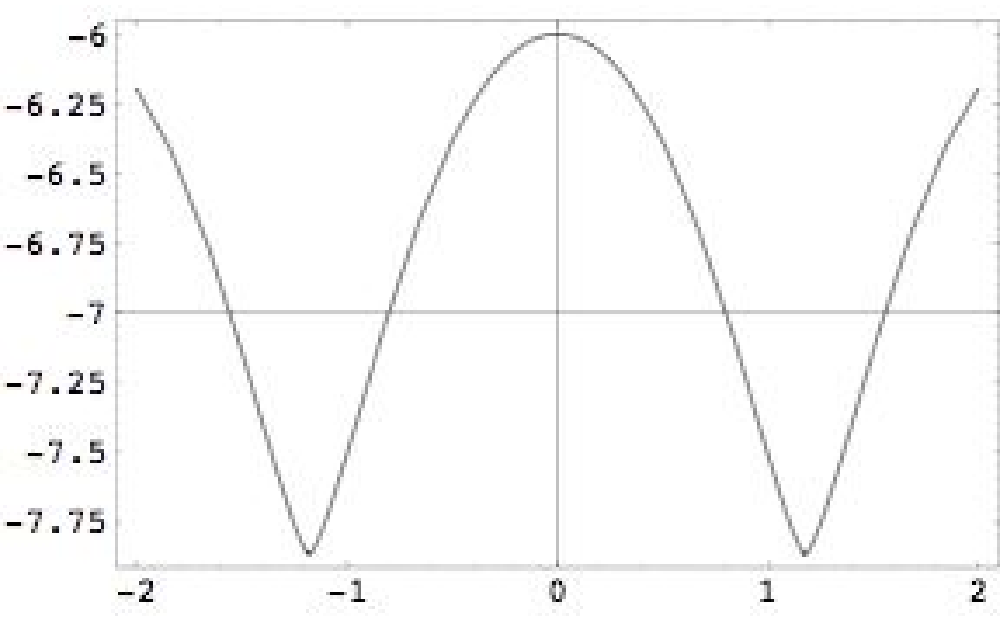}
\end{minipage} \hfill
\begin{minipage}[h]{0.45\linewidth}
\includegraphics[width=5cm]{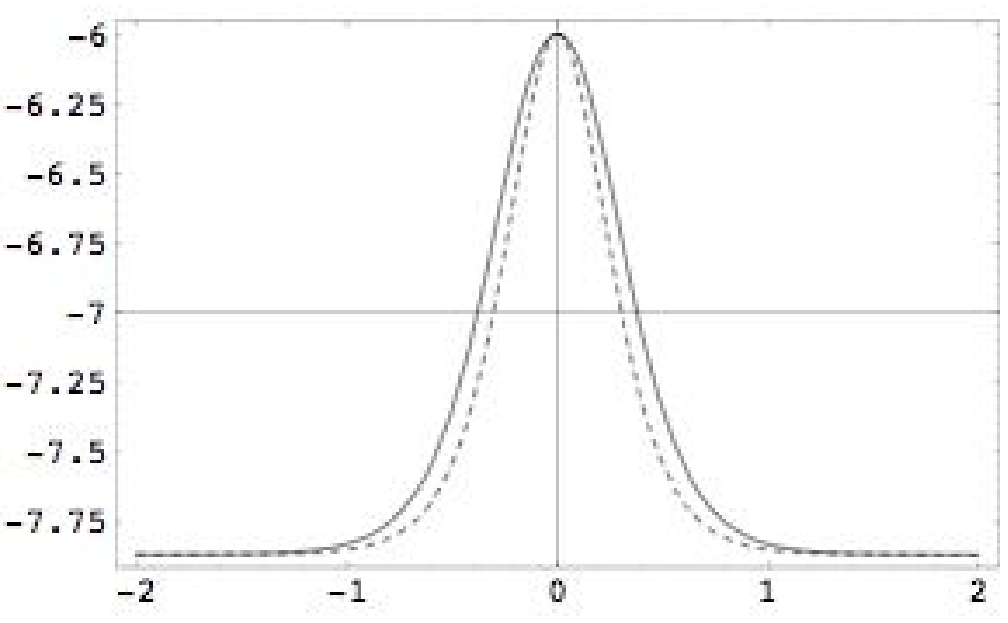}
\end{minipage}
\caption{Plots of the new scalar potential $V(\phi)$ for
$\delta=1/4$ and $\beta=1$ (left). We also plot (right) the
corresponding energy density $\rho(x)$ (continuous line) and
pressure $p(x)$ (dotted line), for same values of $\delta$ and
$\beta$.}\label{potencialGoetz}
\end{figure}

In the Fig. \ref{potencialGoetz}, we plot the scalar potential $\ V(\phi)$, the energy density $\ \rho(x)$ and
the pressure density $\ p(x)$ associated to this system. In spite of the singularity in (\ref{MetricGeneral},
\ref{g2dSIrregular}), we see that $V(\phi)$ is a smooth potential with spontaneously broken symmetry
and the energy and pressure density correspond to a thick domain wall embedded in $AdS_{5}$ spacetime.

Remarkably, the scalar potential is different in the regular and irregular case. Thus, we have found two
domain wall spacetimes with $dS$ expansion compatible with the same scalar field. In \cite{Guerrero:2005aw},
we showed that  (\ref{GeneralSol}), with
$g_{1}$ and $g_{2}$ given by (\ref{f goetz}, \ref{g2dSIrregular}) respectively, defines another domain wall
with interesting properties where it is possible to confine gravity.

\section{Irregular double BPS domain walls}

Let us now consider a symmetric thick domain wall spacetime where
the tensor metric is (\ref{MetricGeneral}) for $\beta=0$, with
 \ini
 g_1(x)=\left[1+(\lambda\:x)^{2s}\right]^{1/2s},\label{metrica
doble}
 \fin
where $\lambda\ $ and $\ s\ $ are real constants with $\ s\ $
an odd integer. This solution was presented and broadly
discussed in \cite{Melfo:2002wd,Castillo-Felisola:2004eg} and represents a two-parameter family of plane
symmetric static double domain wall, being
asymptotically $AdS_5$ with a cosmological constant $-6\
\lambda^2$.

The metric with reciprocal metric factor (\ref{metrica doble}) is
a solution to the coupled Einstein-scalar field equations
(\ref{Einstein-scalar0}-\ref{Einstein-scalar}) with
 \ini
 \phi=\phi_{0}\arctan(\lambda^s\:x^s),\qquad
\phi_{0}=\frac{\sqrt{3(2s-1)}}{s},\label{campo doble}
 \fin
 and
 \inia
 V_1(\phi)&=&\frac{3}{2}\ \lambda^2\left[\tan(\phi/\phi_{0})\right]^{2(1-1/s)}\nonumber\\
&&\left[\cos(\phi/\phi_{0})\right]^{2(2-1/s)}\left[-1+2s-4\tan^2(\phi/\phi_{0})\right],\label{potencial
doble1}
 \fina
where $\phi$ interpolates between two degenerate minima of
$V(\phi)$, $\ \phi(\pm \infty)=\pm\ \phi_0\pi/2$. Similar solutions are
also considered in \cite{Bazeia:2003aw}.

To our knowledge, there are few analytic solutions that host a topological double kink configuration described
 by a single real scalar field \cite{Melfo:2002wd, Bazeia:2003aw}. In this sense, we want to obtain another
double domain wall using the method presented in section II. In concordance with our approach, consider (\ref{GeneralSol})
with
$g_1$ given by (\ref{metrica doble})
 \ini
 g=\lambda\: x\:\left[1+(\lambda\:x)^{2s}\right]^{1/2s}{}_2F_1[l,k,n,-(\lambda\:x)^{2s}],\label{g22}
 \fin
where $\ l=1/(2s),\ k=2l\ $ and $\ n=1+l$. This spacetime is also
solution to the coupled system with the scalar field (\ref{campo
doble}) and
 \inia
 V(\phi)&=&-12\lambda^2\sin^2(\phi/\phi_{0})\:{}_2F_1[l,k,n;-\tan^2(\phi/\phi_{0})]
-6\lambda^2\cos^{2/s}(\phi/\phi_{0})\nonumber\\
&&-\frac{3}{8}\lambda^2\sin^2(2\phi/\phi_{0})\cos^{-2/s}(\phi/\phi_{0})\:{}_2F_1[l,k,n;-\tan^2(\phi/\phi_{0})]^{2}\nonumber\\
&&\left[1-2s+4\tan^2(\phi/\phi_{0})\right],\label{potencial
doble}
\fina
where $\ \phi\ $ takes values $\ \pm\ \phi_{0}\ \pi/2$ at $\ x\rightarrow\pm\infty$, corresponding
to two minima of the potential, and interpolates smoothly between them. On the other hand,  the energy
density of the configuration is given by
\inia
\rho(x)&=&-6\lambda^{2}[1+(\lambda\: x)^{2s}]^{-1/s}\nonumber\\
&&-12\lambda^{2}(\lambda\: x)^{2s}[1+(\lambda\: x)^{2s}]^{-1}\:{}_2F_1[l,k,n,-(\lambda\:x)^{2s}]\nonumber\\
&&-3\lambda^{2}(\lambda\: x)^{2s}[1+(\lambda\: x)^{2s}]^{-2+1/s}[1-2s+(\lambda\: x)^{2s}]\nonumber\\
&&{}_2F_1[l,k,n,-(\lambda\:x)^{2s}]^{2}.\label{densidadDoble}
\fina

\begin {figure}[h]
\begin{minipage}[h]{0.45\linewidth}
\includegraphics[width=5cm,angle=0]{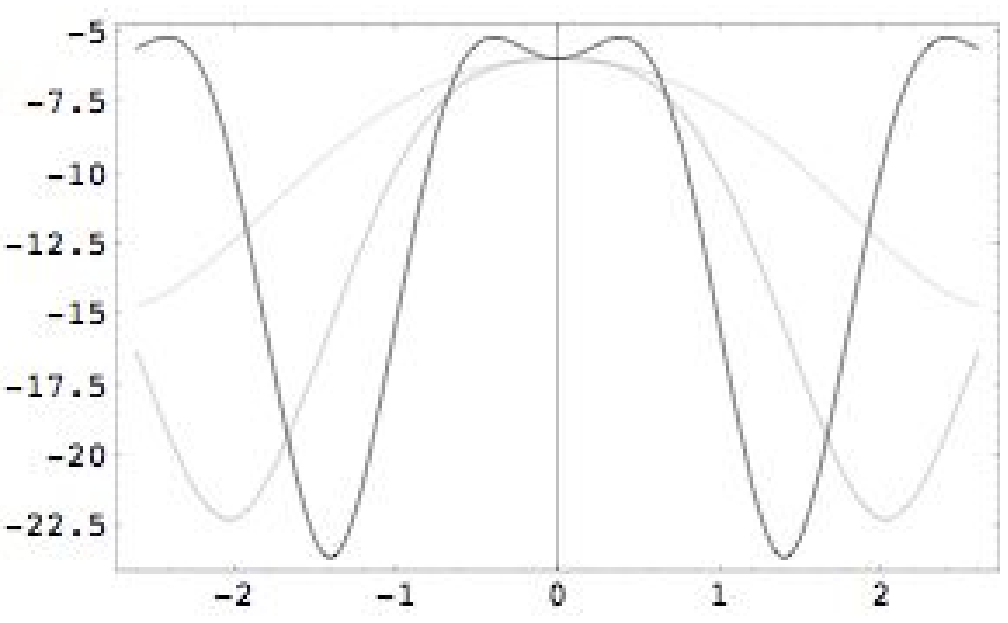}
\end{minipage} \hfill
\begin{minipage}[h]{0.45\linewidth}
\includegraphics[width=5cm]{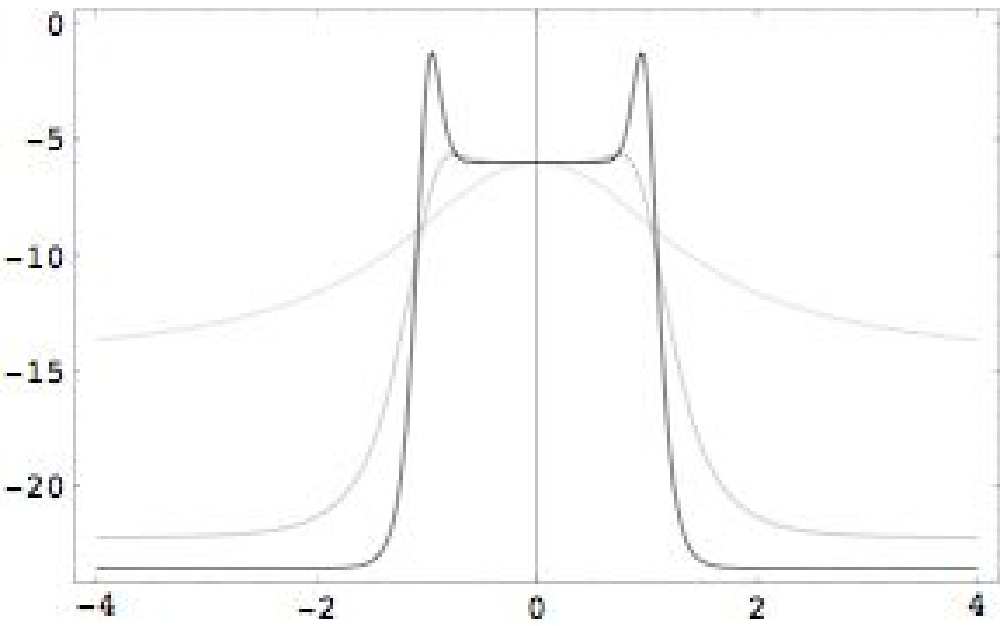}
\end{minipage}
\caption{Graphics of the potential $V(\phi)$ (left) and energy density
$\rho(x)$ (right) for $s=
1, 3$ and $7$. The thickness of the lines
increases with increasing s. }\label{potencialDoble}
\end{figure}

It should be noted that the metric tensor (\ref{MetricGeneral},
\ref{g22}) is a singular tensor across of the hypersurface $\
x=0$. However, this is a two-parameter family of plane symmetric
static double domain wall spacetime with reflection symmetry along
the direction perpendicular to the wall associated to a scalar
potential with spontaneously broken symmetry. These walls
interpolate between $AdS_5$ asymptotic vacua with cosmological
constant $\ -6\lambda^{2}[\ \Gamma(l)\Gamma(n)\ ]^2/\Gamma(k)^{2}\
$. In Fig.\ref{potencialDoble} we plot the potential $\ V(\phi)\
$, and the energy density $\ \rho(x)\ $ for different values of $\
s$.

The energy density shows that the matter field gives rise to thick brane composed of a single ($s=1$)
or two ($s\neq 1$) interfaces. In the last case, one sees the appearance of a new phase in between the two
interfaces where the energy density of the matter field gets more concentrated. As consequence,
a different chart may exist where the metric factor is essentially constant inside the brane
\cite{Melfo:2002wd,Bazeia:2003aw}; but this is of no concern to us here.

Notice that the potential associated to the irregular case is notably different from the regular case.
Thus, we found two solutions to (\ref{Einstein-scalar0}-\ref{Einstein-scalar}) compatible with the scalar field (\ref{campo doble}),
but embedded in different $AdS$ backgrounds. In the Ref. \cite{Guerrero:2005aw}, we considered
the regular and irregular solutions (\ref{metrica doble}, \ref{g22}) in order to construct an asymmetric
brane world with two different walls, where the zero mode of the metric fluctuations is localized on one of them.

Before finishing, we would like to explore the stability of the vacuum. As stated above, the irregular
domain wall spacetime considered in this section is asymptotically $AdS_{5}$. It is widely known
that $AdS$ vacuum stability requires $V(\phi)$ to take the form  \cite{Townsend:1984iu,Boucher:1984yx}
 \ini
 V(\phi)=\frac{9}{2}\left(\frac{dW}{d\phi}\right)^2-6W^2,\label{VyW}
 \fin
where $\ W(\phi)$ is any function with at least one critical point, called fake superpotential \cite{Kallosh:2000tj}. Critical points of $\ W(\phi)$ are also critical points of $V(\phi)$ and in the context of supergravity theories the critical points of $\ W(\phi)$ yield stable $AdS$ vacua \cite{Townsend:1984iu}. 

It follows that in this case
\begin{eqnarray}
W(\phi)&&=\lambda\ |\cos^{1/s}(\phi/\phi_0)|\nonumber\\
&&\left[1+\sin^{2}(\phi/\phi_0)
\cos^{-2/s}(\phi/\phi_0){}_{2}F_{1}[l,k,n;-\tan
^{2}(\phi/\phi_0)]\right],\label{superpotential}
\end{eqnarray}
whose critical points are the asymptotic values of the scalar field, $\ \phi=\pm\pi\phi_{0}/2$. Whether a supergravity theory can be constructed so that the supersymmetry conditions lead to (\ref{superpotential}) is a question beyond the scope of this paper. But since the critical points of (\ref{superpotential}) are asymptotic values of $\phi$, as given by (\ref{potencial doble}), this suggests that these asymptotic $AdS$ vacua are stable. 

Finally, it should be note that (\ref{MetricGeneral}, \ref{g22})
and (\ref{campo doble}, \ref{potencial doble}) can be parametrized by (\ref{superpotential}), so we could have found it using the first order formalism of \cite{Behrndt:1999kz,Skenderis:1999mm,DeWolfe:1999cp}, which in the particular coordinate system (\ref{MetricGeneral}) is given by
\begin{equation}
g^{\prime}=W(\phi),\qquad
g\phi^{\prime}=3\frac{dW(\phi)}{d\phi}.\label{BPSeqs}
\end{equation}
The domain walls obtained from (\ref{VyW}, \ref{BPSeqs}), where the $AdS$ vacua are stable, are solutions that satisfy the Bogomol'nyi-Prasad-Sommerfield (BPS) bound \cite{Bogomolngy,Prasad,Skenderis:1999mm}. In this sense, the static double configuration is a BPS domain wall scenario.

\section{Summary and Remarks}

In this paper we presented a new formulation or method based on
the linearization  of one of the equations of the coupled
Einstein-scalar field system, which gives a mechanism to obtain new
solutions from well known ones. We showed that the geometries
associated to these solutions are connected by a conformal
transformation and, in concordance with \cite{Wald:1984rg}, the
corresponding spacetimes are physically different.

From the approach, we found two thick branes embedded in a
spacetime with a nonconventional geometry. These branes correspond
to domain walls on a topological space $(\mathbb{R}^5,\ {\bf g})$,
which fails to Hausdorff at the origin and turns out to be
singular. However, the components of Riemann tensor in the
explorer's orthonormal frame become bounded in the pathological
point. Thus, we conclude that spacetime geometry has a coordinate
singularity and the hypersurface hosted at the origin is a
well-behaved region of spacetime.

The branes obtained in this paper are explicit realizations of
thick domain wall spacetimes. One of them is a $dS$ expansion
and the other one is a static topological double kink; both being asymptotically
$AdS_5$ with $Z_2$ symmetry along the direction perpendicular to the wall.
These scenarios are the irregular versions of
the domain walls reported in \cite{Goetz:1990,Melfo:2002wd} and we have considered them
in \cite{Guerrero:2005aw} in order to construct exotic brane worlds.

In particular, for the static case, it has been shown that the scalar field potential for this
solution satisfies the requirements for the
existence of stable $AdS$ vacua, being derivable from a fake
superpotential function whose critical points are the asymptotic values of the scalar field. Moreover, this solution can be obtained from the first order formulation of the equations of motion that saturate the BPS bound. Thus, the
family of irregular static topological double kink defines a BPS
domain wall spacetime.

\section*{Acknowledgments}
R.G. and R.O.R. wish to thank  A. Melfo, N. Pantoja and N. Romero
for fruitful discussion and Susana Zoghbi for her collaboration to
complete this paper. This work was supported by CDCHT-UCLA under
project 001-CT-2004.



\begin{thebibliography}{99}

\bibitem{Randall:1999ee}
  L.~Randall and R.~Sundrum,
  Phys.\ Rev.\ Lett.\  {\bf 83}, 3370 (1999)
  [arXiv:hep-ph/9905221].

\bibitem{Randall:1999vf}
  L.~Randall and R.~Sundrum,
  Phys.\ Rev.\ Lett.\  {\bf 83}, 4690 (1999)
  [arXiv:hep-th/9906064].
  
\bibitem{Cvetic:1993xe}
  M.~Cvetic, S.~Griffies and H.~H.~Soleng,
  Phys.\ Rev.\ D {\bf 48}, 2613 (1993)
  [arXiv:gr-qc/9306005].


\bibitem{Maldacena:1997re}
J.~M.~Maldacena,
Adv.\ Theor.\ Math.\ Phys.\  {\bf 2}, 231 (1998)
[Int.\ J.\ Theor.\ Phys.\  {\bf 38}, 1113 (1999)]
[arXiv:hep-th/9711200].



\bibitem{DeWolfe:1999cp}
O.~DeWolfe, D.~Z.~Freedman, S.~S.~Gubser and A.~Karch,
Phys.\ Rev.\ D {\bf 62}, 046008 (2000)
[arXiv:hep-th/9909134].


\bibitem{Gremm:1999pj}
M.~Gremm,
Phys.\ Lett.\ B {\bf 478}, 434 (2000) [arXiv:hep-th/9912060].



\bibitem{Csaki:2000fc}
  C.~Csaki, J.~Erlich, T.~J.~Hollowood and Y.~Shirman,
  Nucl.\ Phys.\ B {\bf 581}, 309 (2000)
  [arXiv:hep-th/0001033].

\bibitem{Kehagias:2000au}
  A.~Kehagias and K.~Tamvakis,
  Phys.\ Lett.\ B {\bf 504}, 38 (2001)
  [arXiv:hep-th/0010112].

\bibitem{Kehagias:2000dg}
  A.~Kehagias and K.~Tamvakis,
  Mod.\ Phys.\ Lett.\ A {\bf 17}, 1767 (2002)
  [arXiv:hep-th/0011006].

\bibitem{Behrndt:2001km}
  K.~Behrndt and G.~Dall'Agata,
  Nucl.\ Phys.\ B {\bf 627}, 357 (2002)
  [arXiv:hep-th/0112136].

\bibitem{Kobayashi:2001jd}
  S.~Kobayashi, K.~Koyama and J.~Soda,
  Phys.\ Rev.\ D {\bf 65}, 064014 (2002)
  [arXiv:hep-th/0107025].

\bibitem{Campos:2001pr}
  A.~Campos,
  Phys.\ Rev.\ Lett.\  {\bf 88}, 141602 (2002)
  [arXiv:hep-th/0111207].




\bibitem{Skenderis:1999mm}
K.~Skenderis and P.~K.~Townsend,
Phys.\ Lett.\ B {\bf 468}, 46 (1999)
[arXiv:hep-th/9909070].

\bibitem{Guerrero:2002ki}
R.~Guerrero, A.~Melfo and N.~Pantoja,
Phys.\ Rev.\ D {\bf 65}, 125010 (2002)
[arXiv:gr-qc/0202011].

\bibitem{Bazeia:2002xg}
D.~Bazeia, L.~Losano and J.~M.~C.~Malbouisson,
Phys.\ Rev.\ D {\bf 66}, 101701 (2002)
[arXiv:hep-th/0209027].

\bibitem{Melfo:2002wd}
A.~Melfo, N.~Pantoja and A.~Skirzewski,
Phys.\ Rev.\ D {\bf 67}, 105003 (2003) [arXiv:gr-qc/0211081].

\bibitem{Guerrero:2005aw}
R.~Guerrero, R.~O.~Rodriguez and R.~Torrealba,
Phys.\ Rev.\ D {\bf 72}, 124012 (2005)
[arXiv:hep-th/0510023].


\bibitem{Goetz:1990} G.~Goetz,
 J.\ Math.\ Phys. {\bf 31}  2683 (1990).

\bibitem{Sasakura:2002tq}
  N.~Sasakura,
  JHEP {\bf 0202}, 026 (2002)
  [arXiv:hep-th/0201130].


\bibitem{Campos:1998db}
  A.~Campos, K.~Holland and U.~J.~Wiese,
  Phys.\ Rev.\ Lett.\  {\bf 81}, 2420 (1998)
  [arXiv:hep-th/9805086].

\bibitem{Morris:1995zp}
  J.~R.~Morris,
  Phys.\ Rev.\ D {\bf 51}, 697 (1995).

\bibitem{Bazeia:1995ui}
  D.~Bazeia, R.~F.~Ribeiro and M.~M.~Santos,
  Phys.\ Rev.\ D {\bf 54}, 1852 (1996).

\bibitem{Edelstein:1997ej}
  J.~D.~Edelstein, M.~L.~Trobo, F.~A.~Brito and D.~Bazeia,
  Phys.\ Rev.\ D {\bf 57}, 7561 (1998)
  [arXiv:hep-th/9707016].

\bibitem{Gregory:2001xu}
  R.~Gregory and A.~Padilla,
  Phys.\ Rev.\ D {\bf 65}, 084013 (2002)
  [arXiv:hep-th/0104262].


\bibitem{Bazeia:2003aw}
  D.~Bazeia, C.~Furtado and A.~R.~Gomes,
  JCAP {\bf 0402}, 002 (2004)
  [arXiv:hep-th/0308034].
  
 \bibitem{Taub}
  A.~H.~Taub,
  Annals Math. {\bf 53}, 472 (1953).
  




\bibitem{Novello:2000km}
  M.~Novello, S.~E.~Perez Bergliaffa and J.~M.~Salim,
  Class.\ Quant.\ Grav.\  {\bf 17}, 3821 (2000)
  [arXiv:gr-qc/0003052].



\bibitem{Wald:1984rg}
R.~M.~Wald,
{\it General Relativity} (University of Chicago Press, Chicago, 1984).


\bibitem{Arfken}
G.~Arfken and H.~Weber,
{\it Mathematical Methods for Physicists} (Academic Press, 1995).


\bibitem{Gass:1999gk}
R.~Gass and M.~Mukherjee,
Phys.\ Rev.\ D {\bf 60}, 065011 (1999)
[arXiv:gr-qc/9903012].

\bibitem{Wang:2002pk}
A.~z.~Wang,
Phys.\ Rev.\ D {\bf 66}, 024024 (2002)
[arXiv:hep-th/0201051].


\bibitem{Pantoja:2002nw}
  N.~R.~Pantoja, H.~Rago and R.~O.~Rodriguez,
  J.\ Math.\ Phys.\  {\bf 45}, 1994 (2004)
  [arXiv:gr-qc/0205094].

\bibitem{Pantoja:1997zt}
  N.~R.~Pantoja and H.~Rago,
  arXiv:gr-qc/9710072.
  
\bibitem{Pantoja:2000cq}
  N.~R.~Pantoja and H.~Rago,
  Int.\ J.\ Mod.\ Phys.\ D {\bf 11}, 1479 (2002)
  [arXiv:gr-qc/0009053].


\bibitem{Castillo-Felisola:2004eg}
  O.~Castillo-Felisola, A.~Melfo, N.~Pantoja and A.~Ramirez,
  Phys.\ Rev.\ D {\bf 70}, 104029 (2004)
  [arXiv:hep-th/0404083].



\bibitem{Boucher:1984yx}
W.~Boucher,
Nucl.\ Phys.\ B {\bf 242}, 282 (1984).

\bibitem{Townsend:1984iu}
P.~K.~Townsend,
Phys.\ Lett.\ B {\bf 148}, 55 (1984).

\bibitem{Kallosh:2000tj}
  R.~Kallosh and A.~D.~Linde,
  JHEP {\bf 0002}, 005 (2000)
  [arXiv:hep-th/0001071].


\bibitem{Behrndt:1999kz}
  K.~Behrndt and M.~Cvetic,
  Phys.\ Lett.\ B {\bf 475}, 253 (2000)
  [arXiv:hep-th/9909058].
  
\bibitem{Bogomolngy}
E.~Bogomol'nyi,
Sov.\ J.\ Nucl.\ Phys. {\bf 24}, 449 (1976).

\bibitem{Prasad} 
M.~K.~Prasad and C.~H.~Sommerfield, 
Phys.\ Rev.\ Lett. {\bf 35}, 760 (1975).



\end{thebibliography}
\end{document}